\documentclass[conference]{IEEEtran}

\IEEEoverridecommandlockouts
\usepackage{cite}
\usepackage{overpic}
\usepackage{amsmath,amssymb,amsfonts}

\usepackage{graphicx}
\usepackage{textcomp}
\usepackage{xcolor}
\usepackage{float}
\usepackage{amsthm}
\usepackage{graphicx}
\usepackage{epstopdf}
\usepackage{amsmath,bm}
\usepackage{amsfonts}
\usepackage{amssymb}
\usepackage{color}
\usepackage{multirow}
\usepackage{multicol}
\usepackage{soul,xcolor}
\usepackage{algorithm}
\usepackage{algpseudocode}%

\usepackage{comment}

\newcommand{\mathacr}[1]{\mathsf{#1}}
\theoremstyle{plain}

\newtheorem{lemma}{Lemma}

\newcommand{\vect}[1]{\mathbf{#1}}

\def\kron{\otimes}
\def\tr{\mathrm{tr}}

\def\Htran{\mbox{\tiny $\mathrm{H}$}}
\def\Ttran{\mbox{\tiny $\mathrm{T}$}}
\def\CN{\mathcal{N}_{\mathbb{C}}} 

\begin{document}

\title{Point-to-Point MIMO Channel Estimation by Exploiting Array Geometry and Clustered Multipath Propagation  \vspace{-0.2cm}
\thanks{The work by \"O. T. Demir was supported by 2232-B International Fellowship for Early Stage Researchers Programme funded by the Scientific and Technological Research Council of T\"urkiye. The work by E. Bj{\"o}rnson was supported by the  FFL18-0277 grant from the Swedish Foundation for Strategic Research.}}

\author{\IEEEauthorblockN{ \"Ozlem Tu\u{g}fe Demir$^*$ and Emil Bj{\"o}rnson$^{\dagger}$}
\IEEEauthorblockA{{$^*$Department of Electrical-Electronics Engineering, TOBB University of Economics and Technology, Ankara, T\"urkiye
		} \\ {$^\dagger$Department of Computer Science, KTH Royal Institute of Technology, Stockholm, Sweden  
		} \\
		{Email: ozlemtugfedemir@etu.edu.tr, emilbjo@kth.se  \vspace{-0.4cm}}
}

}

\maketitle

\begin{abstract}
A large-scale MIMO (multiple-input multiple-output) system offers significant advantages in wireless communication, including potential spatial multiplexing and beamforming capabilities. However, channel estimation becomes challenging with multiple antennas at both the transmitter and receiver ends. The minimum mean-squared error (MMSE) estimator, for instance, requires a spatial correlation matrix whose dimensions scale with the square of the product of the number of antennas on the transmitter and receiver sides. This scaling presents a substantial challenge, particularly as antenna counts increase in line with current technological trends. Traditional MIMO literature offers alternative channel estimators that mitigate the need to fully acquire the spatial correlation matrix. In this paper, we revisit point-to-point MIMO channel estimation and introduce a reduced-subspace least squares (RS-LS) channel estimator designed to eliminate physically impossible channel dimensions inherent in uniform planar arrays. Additionally, we propose a cluster-aware RS-LS estimator that leverages both reduced and cluster-specific subspace properties, significantly enhancing performance over the conventional RS-LS approach. Notably, both proposed methods obviate the need for fully/partial knowledge of the spatial correlation matrix.
\end{abstract}


\section{Introduction}
Equipping base stations (BSs) with a large number of antennas enhances beamforming gains and facilitates spatial multiplexing in single-user and multi-user multiple-input multiple-output (MIMO) setups. In the context of what is widely referred to as \emph{massive MIMO}, it is generally assumed that each user equipment (UE) is equipped with a single antenna. The term MIMO is applicable here because multiple UEs—each with at least one antenna—are simultaneously communicating with a multi-antenna BS.  The true advantage of massive MIMO lies in the precise channel estimation achieved through uplink pilots. Assigning orthogonal pilot sequences to UEs allows for effective channel despread and the design of combining/precoding vectors that enhance the data rate for each UE.

Recent advancements have also reignited interest in what the classical MIMO literature describes as \emph{point-to-point MIMO} \cite{bjornson2024introduction}. This mode is relevant when a single UE with multiple antennas communicates with the BS, allowing multiple data streams per UE \cite{massive-MIMO-multiple-antenna-UEs,cell-free-multi-antenna}. Although only a single UE with multiple antennas might be involved, channel estimation remains a formidable challenge due to the significant increase in the number of parameters—product of the transmit and receive antenna counts.

Past solutions to related problems, such as training signal design for classical MIMO with correlated channels \cite{Kotecha2004a,Liu2007a}, predominantly focused on minimum mean-squared error (MMSE) channel estimation. However, in the context of today's vast antenna arrays, fully characterizing the spatial correlation of MIMO channels is often impractical. A novel approach is presented in \cite{shariati2014low}, where the authors adapt the MMSE estimator to use only the diagonal entries of the spatial correlation matrix of the vectorized MIMO channel, termed the element-wise MMSE estimator.

In this paper, we introduce a channel estimation scheme that outperforms traditional estimators without requiring knowledge of the spatial correlation matrix. We extend the spatial correlation model for uniform linear arrays (ULAs) with azimuthal variations \cite{Sayeed2002a} to uniform planar arrays (UPAs) that include elevation angle variations. Building on our prior work \cite{Demir2021RISb,demir2022exploiting}, we develop the reduced-subspace least-squares (RS-LS) estimator, which calculates a reduced-dimension subspace capturing all possible spatial correlation and channel matrices. We further refine this method by introducing a cluster-aware RS-LS estimator that enhances channel estimation accuracy by utilizing the array geometry and cluster subspace structure. These advancements allow our estimators to surpass traditional LS and MMSE estimators in terms of mean squared error (MSE) performance, while reducing the number of pilots needed for accurate estimation.

\section{System and Channel Modeling}

We examine a point-to-point MIMO communication scenario between a transmitter with $K$ antennas and a receiver with $M$ antennas, both employing a UPA configuration. The transmitter's array consists of $K_{\rm H}$ antennas per row and $K_{\rm V}$ antennas per column, resulting in a total of $K=K_{\rm H}K_{\rm V}$ antennas. Similarly, the receiver's array has $M_{\rm H}$ antennas per row and $M_{\rm V}$ antennas per column, leading to a total of $M=M_{\rm H}M_{\rm V}$ antennas.

At both the transmitter and receiver, the antennas are indexed as shown in \cite[Fig.~4.37]{bjornson2024introduction}. The array response vectors for the transmitter and receiver UPAs are denoted by $\vect{a}_{K_{\rm H},K_{\rm V}}(\varphi,\theta)$ and $\vect{a}_{M_{\rm H},M_{\rm V}}(\varphi,\theta)$, respectively, where $\varphi$ and $\theta$ represent the azimuth and elevation angles \cite[Eqn.~(4.128)]{bjornson2024introduction}.

We consider a time-varying channel, employing a conventional block fading model that divides time/frequency resources into coherence blocks with frequency-flat and static channel realizations. Within each block, $\tau_p$ samples are allocated for pilot transmission. The channel from the transmitter to the receiver array is represented by $\vect{H}\in \mathbb{C}^{M\times K}$, which assumes an independent realization in each coherence block. In UPAs, spatial correlation between elements is inherent, even in the presence of isotropic scattering and isotropic antennas, due to spatial oversampling in the array \cite{Demir2021RISb}. This characteristic implies that the set of physically plausible channel realizations span a lower-dimensional subspace of $\mathbb{C}^{M\times K}$. In this paper, we aim to leverage this subspace to achieve efficient channel estimation without precise knowledge of the spatial statistics of the channel $\vect{H}$.

Extending the general MIMO channel representation in \cite{Sayeed2002a} to the UPA case, a channel $\vect{H}$ can be expressed as
\begin{align}
\vect{H} &= \sqrt{\beta}\iiiint_{-\pi/2}^{\pi/2}g_{\rm{r,t}}(\varphi_{\rm{r}},\theta_{\rm{r}},\varphi_{\rm{t}},\theta_{\rm{t}}) \nonumber\\
&\quad\times\vect{a}_{M_{\rm H},M_{\rm V}}(\varphi_{\rm{r}},\theta_{\rm{r}})\vect{a}_{K_{\rm H},K_{\rm V}}^{\Ttran}(\varphi_{\rm{t}},\theta_{\rm{t}})d\varphi_{\rm{r}}d\theta_{\rm{r}}d\varphi_{\rm{t}}d\theta_{\rm{t}}, \label{eq:H}
\end{align}
where $\beta>0$ is the average channel gain and $g_{\rm{r,t}}(\varphi_{\rm{r}},\theta_{\rm{r}},\varphi_{\rm{t}},\theta_{\rm{t}})$ is the \emph{spatial spreading function} that represents the physical scattering and characterizes the gain and phase-shift of the cluster from its direction $(\varphi_{\rm{t}},\theta_{\rm{t}})$ as observed from the transmitter and $(\varphi_{\rm{r}},\theta_{\rm{r}})$ as observed from the receiver. We note that the waves only arrive/depart to/from the directions in front of the arrays so that $\varphi_{\rm{r}},\varphi_{\rm{t}}\in \left[-\frac{\pi}{2},\frac{\pi}{2}\right]$. Following \cite{Sayeed2002a}, we model $g_{\rm{r,t}}(\varphi_{\rm{r}},\theta_{\rm{r}},\varphi_{\rm{t}},\theta_{\rm{t}})$ as a spatially uncorrelated circularly symmetric Gaussian stochastic process with cross-correlation 
\begin{align}
&\mathbb{E}\left\{g_{\rm{r,t}}(\varphi_{\rm{r}},\theta_{\rm{r}},\varphi_{\rm{t}},\theta_{\rm{t}})g_{\rm{r,t}}^*(\varphi^{\prime}_{\rm{r}},\theta^{\prime}_{\rm{r}},\varphi^{\prime}_{\rm{t}},\theta^{\prime}_{\rm{t}})\right\}\nonumber\\
  &=f_{\rm r,t}(\varphi_{\rm r},\theta_{\rm r},\varphi_{\rm t},\theta_{\rm t})\delta(\varphi_{\rm r}-\varphi^{\prime}_{\rm r})\delta(\theta_{\rm r}-\theta^{\prime}_{\rm r})\nonumber \\
  &\quad \times\delta(\varphi_{\rm t}-\varphi^{\prime}_{\rm t})\delta(\theta_{\rm t}-\theta^{\prime}_{\rm t}),  
\end{align}
where $\delta(\cdot)$ is the Dirac delta function and $f_{\rm r,t}(\varphi_{\rm r},\theta_{\rm r},\varphi_{\rm t},\theta_{\rm t})\geq0$ is the \emph{spatial scattering function}. This function takes the role of a joint probability density function of the azimuth and elevation angles, i.e., 
\begin{align}
   & \iiiint_{-\pi/2}^{\pi/2}f_{\rm r,t}(\varphi_{\rm r},\theta_{\rm r},\varphi_{\rm t},\theta_{\rm t})d\varphi_{\rm r}d\theta_{\rm r}d\varphi_{\rm t}d\theta_{\rm t} = 1.
\end{align}
We define $\vect{x}=\mathrm{vec}\left(\vect{H}^{\Ttran}\right)$ and then, it follows that
\begin{align}
\vect{x}\sim \CN (\vect{0},\vect{R}),
    \end{align}
    which is a vectorized channel being correlated Rayleigh fading with the spatial correlation matrix 
    \begin{align}
        \vect{R}& = \mathbb{E}\left\{\vect{x}\vect{x}^{\Htran}\right\} = \beta \iiiint_{-\pi/2}^{\pi/2}f_{\rm r,t}(\varphi_{\rm r},\theta_{\rm r},\varphi_{\rm t},\theta_{\rm t})\nonumber\\
        &\quad \times\left(\vect{a}_{M_{\rm H},M_{\rm V}}(\varphi_{\rm{r}},\theta_{\rm{r}}) \kron \vect{a}_{K_{\rm H},K_{\rm V}} (\varphi_{\rm{t}},\theta_{\rm{t}})\right)\nonumber\\
        & \hspace{-4mm}\times \left(\vect{a}_{M_{\rm H},M_{\rm V}}(\varphi_{\rm{r}},\theta_{\rm{r}}) \kron \vect{a}_{K_{\rm H},K_{\rm V}} (\varphi_{\rm{t}},\theta_{\rm{t}})\right)^{\Htran}d\varphi_{\rm r}d\theta_{\rm r}d\varphi_{\rm t}d\theta_{\rm t}, \label{eq:spatial-correlation0}
    \end{align}
    where $\tr(\vect{R})=MK\beta$.

\section{Pilot Transmission and Channel Estimation}

To enable efficient MIMO data transmission, in each coherence block, the receiver must estimate the channel $\vect{H}$. To this end, the transmitter sends a predefined pilot sequence $\vect{\Phi}\in \mathbb{C}^{\tau_p \times K}$ during $\tau_p$ channel uses. The $k$th column of the pilot matrix $\vect{\Phi}$ is the pilot sequence transmitted from transmit antenna $k$. The received signal at the receive antenna $m$ during this training phase is 
\begin{equation}
    \vect{y}_m = \sqrt{\rho}\boldsymbol{\Phi}\vect{h}'_m+ \vect{n}_m, \quad m=1,\ldots,M, 
\end{equation}
where $\vect{h}'_{m}\in \mathbb{C}^{K}$ is the transpose of the $m$th row of $\vect{H}$ and $\vect{n}_m\sim \CN\left(\vect{0}, \vect{I}_{\tau_p} \right)$ is the receiver noise vector whose samples are independent between different receiver antennas. The pilot matrix satisfies $\|\vect{\Phi}\|^2_{\rm F}=\tau_p$ and and $\rho>0$ is the pilot signal-to-noise ratio (SNR). Collecting the received signals for all the BS antennas, we obtain
\begin{equation} \label{eq:concat}
   \underbrace{ \begin{bmatrix}\vect{y}_1 \\ \vdots \\ \vect{y}_M \end{bmatrix}}_{\triangleq \vect{y}}= \sqrt{\rho}\underbrace{\left(\vect{I}_M \kron \vect{\Phi}\right)}_{\triangleq \vect{\Phi}_M} \underbrace{ \begin{bmatrix}\vect{h}'_1 \\ \vdots \\ \vect{h}'_M \end{bmatrix}}_{= \vect{x}} + \ \underbrace{ \begin{bmatrix}\vect{n}_1 \\ \vdots \\ \vect{n}_M \end{bmatrix}}_{\triangleq \vect{n}}. 
\end{equation}
 Following \cite{Liu2007a,Kotecha2004a}, if the spatial correlation matrix $\vect{R}$ in \eqref{eq:spatial-correlation0} is known, then the MMSE estimate of the vectorized channel $\vect{x}$ is computed as
\begin{align}
\widehat{\vect{x}}_{\rm MMSE} = \sqrt{\rho}\vect{R}\vect{\Phi}_M^{\Htran}\left(\rho\vect{\Phi}_M\vect{R}\vect{\Phi}_M^{\Htran}+\vect{I}_{M\tau_p}\right)^{-1}\vect{y}.
\end{align}

We assume that the spatial correlation matrix $\mathbf{R}$ is \emph{unknown} at the receiver due to the practical challenge of acquiring $M^2K^2$ coefficients. Under these conditions, the vector ${\bf x}$ can be estimated using alternative methods. The simplest is the LS estimator~\cite{Kay1993a} 
\begin{align}
    \widehat{\vect{x}}_{\rm LS} =\frac{1}{\sqrt{\rho}} (\vect{\Phi}_M^{\Htran}\vect{\Phi}_M)^{-1}\vect{\Phi}_M^{\Htran}\vect{y},
\end{align}
which necessitates $\tau_p = K$. This requirement may become impractical, particularly for extremely large aperture arrays. An alternative method is the element-wise MMSE estimator, which only requires the diagonal elements of $\mathbf{R}$ \cite{shariati2014low}, i.e.,
\begin{align}
\widehat{\vect{x}}_{\rm EW-MMSE} = \sqrt{\rho}\dot{\vect{R}}\vect{\Phi}_M^{\Htran}\left(\rho\vect{\Phi}_M\dot{\vect{R}}\vect{\Phi}_M^{\Htran}+\vect{I}_{M\tau_p}\right)^{-1}\vect{y},
\end{align}
where $\dot{\vect{R}}$ is the diagonal matrix with the diagonal entries of $\vect{R}$ and there is no restriction on $\tau_p$.

  In \cite{Demir2021RISb,demir2022exploiting}, an RS-LS channel estimator is proposed by only exploiting array geometry and it has been shown that it outperforms the conventional LS estimator significantly. In line with that,  we will now outline the RS-LS channel estimator for the point-to-point MIMO channel and the optimal pilot design that can be applied for  $\tau_p<K$.

\subsection{Reduced-Subspace Least Squares Estimation}
  At this stage, let us assume that there exists a reduced subspace spanned by the columns of the semi-unitary matrix $\vect{U}\in \mathbb{C}^{MK \times r}$ in which any plausible channel vector $\vect{x}$ resides. Here, we denote the dimension of the reduced subspace by $r$. Soon, we will show that such a low-dimension subspace exists, and it can be computed from the eigenspace of the  Kronecker product of two spatial correlation matrices corresponding to the isotropic scattering.

The idea of RS-LS channel estimation is that the channel vector $\vect{x}$ can be expressed as $\vect{U}\vect{w}$ where the elements of $\vect{w}$ are estimated using the LS in the reduced subspace, allowing to suppress noise in unused dimensions. The distribution of $\vect{w}$ is unknown. The RS-LS estimate of $\vect{x}$ is obtained as follows:
\begin{enumerate}
\item  Obtain the LS estimate of $\vect{w}$ in the subspace spanned by the columns of $\vect{U}$;
\item  Bring the estimate back to the original $MK$-dimensional space by multiplying the signal by $\vect{U}$. 
\end{enumerate}
Assuming $M\tau_p\geq r$, the RS-LS estimate of $\vect{x}$ takes the form
\begin{align} \label{eq:RS-LS-estimate}
    \widehat{\vect{x}}_{\rm RS-LS} = \frac{1}{{\sqrt{\rho}}}\vect{U}\left(\vect{U}^{\Htran}\vect{\Phi}_M^{\Htran}\vect{\Phi}_M\vect{U}\right)^{-1}\vect{U}^{\Htran}\vect{\Phi}_M^{\Htran}\vect{y}.
\end{align}
The RS-LS estimation is performed in the $r$-dimensional reduced-subspace and enables the removal of the noise from the non-existing dimensions. The following lemma describes the recipe how to find $\vect{U}$.

\begin{lemma} \label{lemma:span} Let $\overline{\vect{R}}$ and $\vect{R}$ be two spatial correlation matrices for the vectorized channel $\vect{x}$ obtained using the same  transmitter and receiver array geometry. We let the spatial scattering functions corresponding to the correlation matrices $\overline{\vect{R}}$ and $\vect{R}$ according to the correlated fading model in \eqref{eq:spatial-correlation0} be denoted by $\overline{f}_{\rm r,t}(\varphi_{\rm r},\theta_{\rm r},\varphi_{\rm t},\theta_{\rm t})$ and $f_{\rm r,t}(\varphi_{\rm r},\theta_{\rm r},\varphi_{\rm t},\theta_{\rm t})$, respectively, for $\varphi_{\rm r},\varphi_{\rm t}\in[-\pi/2,\pi/2]$ and $\theta_{\rm r},\theta_{\rm t}\in[-\pi/2,\pi/2]$. We assume that the spatial scattering functions are either continuous at each point on its domain or contain Dirac delta functions.

If the domain of  $\overline{f}_{\rm r,t}(\varphi_{\rm r},\theta_{\rm r},\varphi_{\rm t},\theta_{\rm t})$ for which $\overline{f}_{\rm r,t}(\varphi_{\rm r},\theta_{\rm r},\varphi_{\rm t},\theta_{\rm t})>0$ contains the domain of  $f_{\rm r,t}(\varphi_{\rm r},\theta_{\rm r},\varphi_{\rm t},\theta_{\rm t})$ for which $f_{\rm r,t}(\varphi_{\rm r},\theta_{\rm r},\varphi_{\rm t},\theta_{\rm t})>0$, then the subspace spanned by the columns of $\overline{\vect{R}}$ contains the subspace spanned by the columns of $\vect{R}$. 
\end{lemma}

\begin{IEEEproof}
The proof extends the proof of \cite[Lem.~3]{Demir2021RISb} by using  spatial scattering functions $f_{\rm r,t}(\varphi_{\rm r},\theta_{\rm r},\varphi_{\rm t},\theta_{\rm t})$ and array response vectors $\left(\vect{a}_{M_{\rm H},M_{\rm V}}(\varphi_{\rm{r}},\theta_{\rm{r}}) \kron \vect{a}_{K_{\rm H},K_{\rm V}} (\varphi_{\rm{t}},\theta_{\rm{t}})\right)$ for the UPAs in terms of $(\varphi_{\rm r},\theta_{\rm r},\varphi_{\rm t},\theta_{\rm t})$ on the four-dimensional angular domain. 
\end{IEEEproof}

According to Lemma~1, as long as we select a spatial scattering function such that $\mathcal{F}\subset \overline{\mathcal{F}}$, where $\{(\varphi_{\rm r},\theta_{\rm r},\varphi_{\rm t},\theta_{\rm t})\in\mathcal{F}:f_{\rm r,t}(\varphi_{\rm r},\theta_{\rm r},\varphi_{\rm t},\theta_{\rm t})>0\}$ and $\{(\varphi_{\rm r},\theta_{\rm r},\varphi_{\rm t},\theta_{\rm t})\in\overline{\mathcal{F}}:\overline{f}_{\rm r,t}(\varphi_{\rm r},\theta_{\rm r},\varphi_{\rm t},\theta_{\rm t})>0\}$, we can guarantee that the eigenspace of the respective $\overline{\vect{R}}$ spans all plausible channel realizations. If $f_{\rm r,t}(\varphi_{\rm r},\theta_{\rm r},\varphi_{\rm t},\theta_{\rm t})>0$ on $\varphi_{\rm r},\varphi_{\rm t}\in \left[-\frac{\pi}{2},\frac{\pi}{2}\right]$ and $\theta_{\rm r},\theta_{\rm t}\in\left(-\frac{\pi}{2},\frac{\pi}{2}\right)$, one possible selection of $f_{\rm r,t}(\varphi_{\rm r},\theta_{\rm r},\varphi_{\rm t},\theta_{\rm t})=\cos(\theta_{\rm r})\cos(\theta_{\rm t})/(4\pi^2)$, which is the multiplication of two spatial scattering functions corresponding to isotropic scattering with isotropic antennas. Then, $\overline{\vect{R}}$ becomes as shown in \eqref{eq:overlineR} at the top of the next page,
\begin{figure*}
\begin{align}
    \overline{\vect{R}}& = \beta\underbrace{\left( \iint_{-\pi/2}^{\pi/2}\frac{\cos(\theta_{\rm r})}{2\pi}\vect{a}_{M_{\rm H},M_{\rm V}}(\varphi_{\rm r},\theta_{\rm r})\vect{a}_{M_{\rm H},M_{\rm V}}^{\Htran}(\varphi_{\rm r},\theta_{\rm r})d\varphi_{\rm r}d\theta_{\rm r}\right)}_{=\vect{R}_{\rm r,iso}} \kron \underbrace{\left( \iint_{-\pi/2}^{\pi/2}\frac{\cos(\theta_{\rm t})}{2\pi}\vect{a}_{K_{\rm H},K_{\rm V}}(\varphi_{\rm t},\theta_{\rm t})\vect{a}^{\Htran}_{K_{\rm H},K_{\rm V}}(\varphi_{\rm t},\theta_{\rm t})d\varphi_{\rm t}d\theta_{\rm t}\right)}_{=\vect{R}_{\rm t,iso}}. \label{eq:overlineR}
\end{align}
\hrulefill
\end{figure*}
where we denote the resulting normalized correlation matrices at the receiver side by $\vect{R}_{\rm r,iso}$ and at the transmitter side by $\vect{R}_{\rm t,iso}$. The closed-form expressions for the entries of these matrices are given in \cite{Demir2021RISb}. Following Lemma~\ref{lemma:span}, when we use the eigenspace of  $\overline{\vect{R}}=\vect{R}_{\rm r, iso}\kron\vect{R}_{\rm t, iso}$  in the RS-LS estimator, 
we ensure that all \emph{plausible} channel subspaces are included. We denote $\vect{U}=\vect{U}_{\rm r} \kron \vect{U}_{\rm t}\in \mathbb{C}^{MK\times r}$ as the semi-unitary matrix with the columns being orthonormal eigenvectors corresponding to the  $r$ non-zero eigenvalues of $\overline{\vect{R}}$.  Hence, any $\vect{x}$ can be expressed as $\vect{U}\vect{w}$ 
for some reduced-dimension vector $\vect{w} \in \mathbb{C}^{r}$.

The proposed RS-LS estimator can be written from \eqref{eq:RS-LS-estimate} as
\begin{align} \label{eq:RS-LS-estimate-approx}
    &\widehat{\vect{x}}_{\rm RS-LS} = \frac{1}{\sqrt{\rho}}\vect{U}\left(\vect{U}^{\Htran}\vect{\Phi}_M^{\Htran}\vect{\Phi}_M\vect{U}\right)^{-1}\vect{U}^{\Htran}\vect{\Phi}_M^{\Htran}\vect{y} \nonumber\\
    &= \underbrace{\vect{U}\vect{w}}_{=\vect{x}}+\frac{1}{\sqrt{\rho}}\vect{U}\left(\vect{U}^{\Htran}\vect{\Phi}_M^{\Htran}\vect{\Phi}_M\vect{U}\right)^{-1}\vect{U}^{\Htran}\vect{\Phi}_M^{\Htran}\vect{n}
\end{align}
where we assume $r_{\rm t}\leq \tau_p\leq  K$ and $r_{\rm t}$ is the rank of $\vect{R}_{\rm t,iso}$. The respective MSE is
\begin{align} \label{eq:mse-rsls}
    {\rm MSE}_{\rm RS-LS}&= \frac{1}{{\rho}}\tr\left(\vect{U}\left(\vect{U}^{\Htran}\vect{\Phi}_M^{\Htran}\vect{\Phi}_M\vect{U}\right)^{-1}\vect{U}^{\Htran}\right)\nonumber \\
    &=\frac{1}{{\rho}}\tr\left(\left(\vect{U}^{\Htran}\vect{\Phi}_M^{\Htran}\vect{\Phi}_M\vect{U}\right)^{-1}\right)
\end{align}
and depends on the pilot matrix $\vect{\Phi}$. Following the approach \cite{demir2022exploiting}, ${\rm MSE}_{\rm RS-LS}$ can be minimized by the optimal pilot matrix 
\begin{align} \label{eq:optimal-phase-shift}
    \vect{\Phi}^{\star} = \sqrt{\frac{\tau_p}{r_{\rm t}}}\vect{S}_{{\rm \Phi}}\vect{U}_{{\rm t}}^{\Htran}
\end{align}
where $\vect{S}_{{\rm \Phi}}\in \mathbb{C}^{\tau_p\times r_{\rm t}}$ is an arbitrary matrix with orthonormal columns, such as $r_{\rm t}$ columns of a $\tau_p$-dimensional discrete Fourier transform (DFT) matrix. The pilot matrix given above both satisfies $\|\vect{\Phi}^{\star}\|_{\rm F}^2=\tau_p$ and is the global optimal solution for the minimization of the MSE of the RS-LS estimator.

\section{Exploiting Clustered Multipath Propagation}

 In this section, we will utilize a specific structure of the clustered multipath propagation (please see \cite[Fig.~5.29]{bjornson2024introduction}). Assume that the spatial statistics regarding $\vect{H}$ is still not fully known. However, it is known that the vectorized channel can be expressed as
 \begin{align}
    \vect{x} = \sum_{\ell=1}^L \underbrace{\left(\vect{U}_{{\rm r},\ell} \kron 
    \vect{U}_{{\rm t},\ell}\right)\vect{w}_{\ell}}_{=\vect{x}_{\ell}} \label{eq:clustered-channel}
\end{align}
where $\vect{w}_{\ell}$ and $\vect{w}_{\ell^{\prime}}$ are assumed to be independent for $\ell\neq \ell^{\prime}$. Here the clusters are partitioned into $L$ groups such that the corresponding transmit subspaces are mutually orthogonal. To denote these subspaces, we partition $\vect{U}_{\rm t}$ into $L$ non-overlapping sets of columns, where set $\ell$ consists of $r_{{\rm t},\ell}$ orthonormal columns that are put together into the matrix portion $\vect{U}_{{\rm t},\ell}$. Due to clustered multipath propagation environment, we assume that when the $\ell$th portion of the channel $\vect{x}$, which is illuminated by the respective $\vect{U}_{{\rm t},\ell}$ only arrives to the receiver through the subspace spanned by the orthonormal columns of the matrix $\vect{U}_{{\rm r},\ell}\in \mathbb{C}^{M\times r_{{\rm r},\ell}}$, where $r_{{\rm r,\ell}}\leq r_{\rm r}$. There is no restriction regarding the overlapping columns of $\vect{U}_{{\rm r},\ell}$. For example, $\vect{U}_{{\rm r},\ell}$ and $\vect{U}_{{\rm r},\ell^{\prime}}$ can have common columns for some $\ell \neq \ell^{\prime}$. In this section, we will develop a pilot transmission scheme and RS-LS estimation for the case these subspaces are known and exploited, which we call \emph{cluster-aware RS-LS estimator}.

In the proposed pilot training scheme, we decompose the Gram matrix of the pilot matrix $\vect{\Phi}$ as 
\begin{align}
    \vect{\Phi}^{\Htran} \vect{\Phi} = \sum_{\ell=1}^L \widetilde{\vect{\Phi}}_{\ell},
\end{align}
where the columns of the positive semi-definite matrix $\widetilde{\vect{\Phi}}_{\ell}\in \mathbb{C}^{K\times K}$ lies in the column space of $\vect{U}_{{\rm t},\ell}$ so it holds that
\begin{align}\label{eq:orthogonality}
&\vect{U}_{{\rm t},i}^{\Htran}\widetilde{\vect{\Phi}}_{\ell}=\vect{0}, \quad \forall i\neq \ell.
\end{align}

Defining $\vect{U}_{\ell}=\vect{U}_{{\rm r},\ell} \kron 
    \vect{U}_{{\rm t},\ell}$, the proposed RS-LS estimate of $\vect{x}_{\ell}$ is given as
\begin{align} \label{eq:RS-LS-estimate-approx2}
    \widehat{\vect{x}}_{\ell} = &\frac{1}{\sqrt{\rho}}\vect{U}_{\ell}\left(\vect{U}_{\ell}^{\Htran}\vect{\Phi}_M^{\Htran}\vect{\Phi}_M\vect{U}_{\ell}\right)^{-1}\vect{U}_{\ell}^{\Htran}\vect{\Phi}_M^{\Htran}\vect{y} \nonumber \\
    =& \frac{1}{\sqrt{\rho}}\vect{U}_{\ell}\left(\vect{U}_{\ell}^{\Htran}\left(\vect{I}_M\kron\sum_{i=1}^L\widetilde{\vect{\Phi}}_{i}\right)\vect{U}_{\ell}\right)^{-1}\vect{U}_{\ell}^{\Htran}\vect{\Phi}_M^{\Htran}\vect{y},
    \end{align}
   where we utilized \eqref{eq:RS-LS-estimate} for each subspace portion. The matrix inside the inverse operation can be simplified using the properties of the Kronecker product and the orthogonality condition given in \eqref{eq:orthogonality} as
  \begin{align}
&\vect{U}_{\ell}^{\Htran}\left(\vect{I}_M\kron\sum_{i=1}^L\widetilde{\vect{\Phi}}_{i}\right)\vect{U}_{\ell}   \nonumber\\
&= \left(\vect{U}^{\Htran}_{{\rm r},\ell} \kron \vect{U}^{\Htran}_{{\rm t},\ell}\right)    \left(\vect{I}_M\kron\sum_{i=1}^L\widetilde{\vect{\Phi}}_{i}\right)    \left(\vect{U}_{{\rm r},\ell} \kron 
    \vect{U}_{{\rm t},\ell}\right) \nonumber\\
    &= \underbrace{\left(\vect{U}^{\Htran}_{{\rm r},\ell}\vect{U}_{{\rm r},\ell}\right)}_{=\vect{I}_{r_{{\rm r},\ell}}} \kron \left(\sum_{i=1}^L\vect{U}^{\Htran}_{{\rm t},\ell}\widetilde{\vect{\Phi}}_{i}\vect{U}_{{\rm t},\ell}\right) \nonumber\\
    &= \vect{I}_{r_{{\rm r},\ell}} \kron\left(\vect{U}^{\Htran}_{{\rm t},\ell}\widetilde{\vect{\Phi}}_{\ell}\vect{U}_{{\rm t},\ell}\right), \label{eq:simplified1}
  \end{align}
Inserting the above expression, the received pilot signal $\vect{y}$ in \eqref{eq:concat}, and $\vect{x}$ in \eqref{eq:clustered-channel} into the RS-LS estimate in \eqref{eq:RS-LS-estimate-approx2}, we obtain the expression given in 
\eqref{eq:RS-LS-estimate-approx3} at the top of the next page.

    \begin{figure*}
    \begin{align}\label{eq:RS-LS-estimate-approx3}
\widehat{\vect{x}}_{\ell}=&  \frac{1}{\sqrt{\rho}}\left(\vect{U}_{{\rm r},\ell} \kron 
    \vect{U}_{{\rm t},\ell}\right)\left( \vect{I}_{r_{{\rm r},\ell}} \kron\left(\vect{U}^{\Htran}_{{\rm t},\ell}\widetilde{\vect{\Phi}}_{\ell}\vect{U}_{{\rm t},\ell}\right)^{-1}\right)
\left(\vect{U}^{\Htran}_{{\rm r},\ell} \kron 
    \vect{U}^{\Htran}_{{\rm t},\ell}\right)\left(\vect{I}_M\kron\vect{\Phi}^{\Htran}\right)\underbrace{\left(\sqrt{\rho}\left(\vect{I}_M\kron\vect{\Phi}\right)\sum_{i=1}^L \underbrace{\left(\vect{U}_{{\rm r},i} \kron 
    \vect{U}_{{\rm t},i}\right)\vect{w}_{i}}_{=\vect{x}_{i}}+\vect{n}\right)}_{=\vect{y}} \nonumber\\
    =& \sum_{i=1}^L\left(\left(\vect{U}_{{\rm r},\ell}\vect{U}^{\Htran}_{{\rm r},\ell} \vect{U}_{{\rm r},i}\right)\kron\left( \vect{U}_{{\rm t},\ell}\left(\vect{U}^{\Htran}_{{\rm t},\ell}\widetilde{\vect{\Phi}}_{\ell}\vect{U}_{{\rm t},\ell}\right)^{-1}\underbrace{\vect{U}^{\Htran}_{{\rm t},\ell}\sum_{j=1}^L\widetilde{\vect{\Phi}}_{j}\vect{U}_{{\rm t},i}}_{=\begin{cases}\vect{U}^{\Htran}_{{\rm t},\ell}\widetilde{\vect{\Phi}}_{\ell}\vect{U}_{{\rm t},\ell} & i=\ell\\ \vect{0} & i\neq \ell\end{cases}}  \right)\right)\vect{w}_i \nonumber\\
&+\underbrace{\frac{1}{\sqrt{\rho}}\left(\left(\vect{U}_{{\rm r},\ell}\vect{U}^{\Htran}_{{\rm r},\ell}\right)\kron \left( \vect{U}_{{\rm t},\ell}\left(\vect{U}^{\Htran}_{{\rm t},\ell}\widetilde{\vect{\Phi}}_{\ell}\vect{U}_{{\rm t},\ell}\right)^{-1}\vect{U}^{\Htran}_{{\rm t},\ell}\vect{\Phi}^{\Htran}\right)\right)\vect{n}}_{\triangleq \vect{n}_{\ell}} =\underbrace{\left(\vect{U}_{{\rm r},\ell} \kron \vect{U}_{{\rm t},\ell}\right)\vect{w}_{\ell}}_{=\vect{x}_{\ell}}+\vect{n}_{\ell}=\vect{x}_{\ell}+\vect{n}_{\ell}.
\end{align}
\hrulefill
\end{figure*}
The channel estimate of the overall channel is obtained as
\begin{align}
\widehat{\vect{x}} = \sum_{\ell=1}^L \widehat{\vect{x}}_{\ell} = \sum_{\ell=1}^L \vect{x}_{\ell}+\underbrace{\sum_{\ell=1}^L\vect{n}_{\ell}}_{\widetilde{\vect{n}}} =\vect{x}+\widetilde{\vect{n}}
\end{align}
with the respective MSE of the cluster-aware RS-LS being written as
\begin{align} \label{eq:mse-rsls}
    {\rm MSE}_{\rm RS-LS}^{\rm clustered}&= \tr\left(\mathbb{E}\{\widetilde{\vect{n}}\widetilde{\vect{n}}^{\Htran}\}\right)
\end{align}
which can be simplified as shown in \eqref{eq:mse-rsls2} at the next page.
  \begin{figure*}
    \begin{align}\label{eq:mse-rsls2}
 &   {\rm MSE}_{\rm RS-LS}^{\rm clustered} = \sum_{\ell=1}^L\tr\left(\mathbb{E}\left\{\vect{n}_{\ell}\vect{n}_{\ell}^{\Htran}\right\}\right)+\sum_{\ell=1}^L\sum_{i=1,i\neq \ell}^L\tr\left(\mathbb{E}\left\{\vect{n}_{\ell}\vect{n}_i^{\Htran}\right\}\right) \nonumber\\
    =&\frac{1}{\rho} \sum_{\ell=1}^L\tr\left( \left(\left(\vect{U}_{{\rm r},\ell}\vect{U}^{\Htran}_{{\rm r},\ell}\right)\kron \left( \vect{U}_{{\rm t},\ell}\left(\vect{U}^{\Htran}_{{\rm t},\ell}\widetilde{\vect{\Phi}}_{\ell}\vect{U}_{{\rm t},\ell}\right)^{-1}\vect{U}^{\Htran}_{{\rm t},\ell}\vect{\Phi}^{\Htran}\right)\right)\left(\left(\vect{U}_{{\rm r},\ell}\vect{U}^{\Htran}_{{\rm r},\ell}\right)\kron \left( \vect{\Phi}\vect{U}_{{\rm t},\ell}\left(\vect{U}^{\Htran}_{{\rm t},\ell}\widetilde{\vect{\Phi}}_{\ell}\vect{U}_{{\rm t},\ell}\right)^{-1}\vect{U}_{{\rm t},\ell}^{\Htran}\right)\right)  \right) \nonumber \\
    &+\frac{1}{\rho} \sum_{\ell=1}^L\sum_{i=1,i\neq \ell}^L\tr\left( \left(\left(\vect{U}_{{\rm r},\ell}\vect{U}^{\Htran}_{{\rm r},\ell}\right)\kron \left( \vect{U}_{{\rm t},\ell}\left(\vect{U}^{\Htran}_{{\rm t},\ell}\widetilde{\vect{\Phi}}_{\ell}\vect{U}_{{\rm t},\ell}\right)^{-1}\vect{U}^{\Htran}_{{\rm t},\ell}\vect{\Phi}^{\Htran}\right)\right)\left(\left(\vect{U}_{{\rm r},i}\vect{U}^{\Htran}_{{\rm r},i}\right)\kron \left( \vect{\Phi}\vect{U}_{{\rm t},i}\left(\vect{U}^{\Htran}_{{\rm t},i}\widetilde{\vect{\Phi}}_i\vect{U}_{{\rm t},i}\right)^{-1}\vect{U}_{{\rm t},i}^{\Htran}\right)\right)  \right) \nonumber \\
    =&\frac{1}{\rho} \sum_{\ell=1}^L\tr\left( \left(\left(\vect{U}_{{\rm r},\ell}\vect{U}^{\Htran}_{{\rm r},\ell}\right)\kron \left( \vect{U}_{{\rm t},\ell}\left(\vect{U}^{\Htran}_{{\rm t},\ell}\widetilde{\vect{\Phi}}_{\ell}\vect{U}_{{\rm t},\ell}\right)^{-1}\underbrace{\vect{U}^{\Htran}_{{\rm t},\ell}\vect{\Phi}^{\Htran} \vect{\Phi}\vect{U}_{{\rm t},\ell}}_{=\vect{U}^{\Htran}_{{\rm t},\ell}\widetilde{\vect{\Phi}}_{\ell}\vect{U}_{{\rm t},\ell}}\left(\vect{U}^{\Htran}_{{\rm t},\ell}\widetilde{\vect{\Phi}}_{\ell}\vect{U}_{{\rm t},\ell}\right)^{-1}\vect{U}_{{\rm t},\ell}^{\Htran}\right)\right)  \right) \nonumber \\
    &+\frac{1}{\rho} \sum_{\ell=1}^L\sum_{i=1,i\neq \ell}^L\tr\left( \left(\left(\vect{U}_{{\rm r},\ell}\vect{U}^{\Htran}_{{\rm r},\ell}\vect{U}_{{\rm r},i}\vect{U}^{\Htran}_{{\rm r},i}\right)\kron \left( \vect{U}_{{\rm t},\ell}\left(\vect{U}^{\Htran}_{{\rm t},\ell}\widetilde{\vect{\Phi}}_{\ell}\vect{U}_{{\rm t},\ell}\right)^{-1}\underbrace{\vect{U}^{\Htran}_{{\rm t},\ell}\vect{\Phi}^{\Htran}\vect{\Phi}\vect{U}_{{\rm t},i}}_{=\vect{0}}\left(\vect{U}^{\Htran}_{{\rm t},i}\widetilde{\vect{\Phi}}_i\vect{U}_{{\rm t},i}\right)^{-1}\vect{U}_{{\rm t},i}^{\Htran}\right)\right)  \right) \nonumber \\
    =&\frac{1}{\rho} \sum_{\ell=1}^L\tr\left(\left(\vect{U}_{{\rm r},\ell}\vect{U}^{\Htran}_{{\rm r},\ell}\right)\kron \left( \vect{U}_{{\rm t},\ell}\left(\vect{U}^{\Htran}_{{\rm t},\ell}\widetilde{\vect{\Phi}}_{\ell}\vect{U}_{{\rm t},\ell}\right)^{-1}\vect{U}_{{\rm t},\ell}^{\Htran}\right)\right)   = \frac{1}{\rho}\sum_{\ell=1}^Lr_{{\rm r},\ell}\tr\left(\left(\vect{U}^{\Htran}_{{\rm t},\ell}\widetilde{\vect{\Phi}}_{\ell}\vect{U}_{{\rm t},\ell}\right)^{-1}\right).
    \end{align}
    \hrulefill
    \end{figure*}

Based on the derived MSE, the optimal phase-shift matrix $\vect{\Phi}$ that minimizes the MSE can be found by solving the problem
\begin{subequations}
\begin{align}
& \underset{\{\widetilde{\vect{\Phi}}_{\ell}\}}{\mathacr{minimize}} \ \ \sum_{\ell=1}^Lr_{{\rm r},\ell}\tr\left(\left(\vect{U}^{\Htran}_{{\rm t},\ell}\widetilde{\vect{\Phi}}_{\ell}\vect{U}_{{\rm t},\ell}\right)^{-1}\right)  \label{eq:rsls-opt} \\
&\mathacr{subject \ to} \quad \sum_{\ell=1}^L\tr\left(\widetilde{\vect{\Phi}}_{\ell}\right) = \tau_p.
\end{align}
\end{subequations}
Similar to \eqref{eq:optimal-phase-shift}, ${\rm MSE}_{\rm RS-LS}$ is minimized by 
\begin{align}
    \widetilde{\vect{\Phi}}_{\ell} = s_{\ell} \vect{U}_{{\rm t},\ell}\vect{U}_{{\rm t},\ell}^{\Htran},
\end{align}
where $s_{\ell}>0$ is the scaling parameter that needs to be optimized as
\begin{subequations}
\begin{align}
& \underset{\{s_{\ell}\}}{\mathacr{minimize}} \ \ \sum_{\ell=1}^L\frac{r_{{\rm r},\ell}r_{{\rm t},\ell}}{s_{\ell}}  \label{eq:rsls-opt2} \\
&\mathacr{subject \ to} \quad \sum_{\ell=1}^L r_{{\rm t},\ell} s_{\ell} = \tau_p.
\end{align}
\end{subequations}
From the Karush-Kuhn-Tucker conditions, the optimal $s_{\ell}^{\star}$ is obtained as
\begin{align}
    s_{\ell}^{\star} = \frac{\tau_p\sqrt{r_{{\rm r},\ell}}}{\sum_{i=1}^L r_{{\rm t},i}\sqrt{r_{{\rm r},i}}}.
\end{align}
Hence, the optimal Gram matrix of the pilot matrix becomes
\begin{align}
    \left(\vect{\Phi}^{\star}\right)^{\Htran}\vect{\Phi}^{\star} =\sum_{\ell=1}^L \frac{\tau_p\sqrt{r_{{\rm r},\ell}}}{\sum_{i=1}^L r_{{\rm t},i}\sqrt{r_{{\rm r},i}}}\vect{U}_{{\rm t},\ell}\vect{U}_{{\rm t},\ell}^{\Htran}
\end{align}
from which the optimal pilot matrix can be extracted.

\section{Numerical Results}
In this section, we assess the performance of the proposed channel estimators by measuring their normalized mean squared error (NMSE). In addition to the proposed RS-LS and cluster-aware RS-LS estimators, we include a comparison with the element-wise MMSE estimator described in \cite{shariati2014low}.

We consider a system where both the transmitter and receiver have  $8 \times 8$ UPAs with $K=M=64$. The vertical and horizontal antenna spacing are both quarter-of-the-wavelength. At the transmitter side, the entire angular region is segmented into $L=4$ clusters, with each cluster spanning specific angular ranges: i) $\varphi_{\rm t} \in [-\pi/2, -\pi/6]$, $\theta_{\rm t} \in [-\pi/2, 0]$; ii) $\varphi_{\rm t} \in [-\pi/6, \pi/6]$, $\theta_{\rm t} \in [-\pi/2, 0]$; iii) $\varphi_{\rm t} \in [\pi/6, \pi/2]$, $\theta_{\rm t} \in [-\pi/2, 0]$; iv) $\varphi_{\rm t} \in [-\pi/2, \pi/2]$, $\theta_{\rm t} \in [0, \pi/2]$. To ensure orthogonality between these subspaces, orthogonal eigenspaces are formed by projecting each onto the null-space of the others in sequence. As a result, the transmit subspaces $\mathbf{U}_{{\rm t}, \ell}$, for $\ell=1,\ldots,4$, are orthogonal to each other and collectively cover the entire angular region.

On the receiver side, the eigenspaces $\vect{U}_{{\rm r}, \ell}$ are constructed to account for uniformly distributed scattering across four angular subregions: i) $\varphi_{\rm r} \in [-\pi/2, -\pi/3]$, $\theta_{\rm r} \in [-\pi/2, -\pi/3]$; ii) $\varphi_{\rm r} \in [-\pi/3, 0]$, $\theta_{\rm r} \in [-\pi/3, 0]$; iii) $\varphi_{\rm r} \in [0, \pi/3]$, $\theta_{\rm r} \in [0, \pi/3]$; iv) $\varphi_{\rm r} \in [\pi/3, \pi/2]$, $\theta_{\rm r} \in [\pi/3, \pi/2]$. Collectively, these subspaces encompass the entire azimuth and elevation angular domain. These subspaces will be exploited by the cluster-aware RS-LS estimator to boost the performance of the conventional RS-LS estimator.

\begin{figure}[t!]
\hspace{0mm}
\includegraphics[trim={0.7cm 0.1cm 1.5cm 0.2cm},clip,width=3.2in]{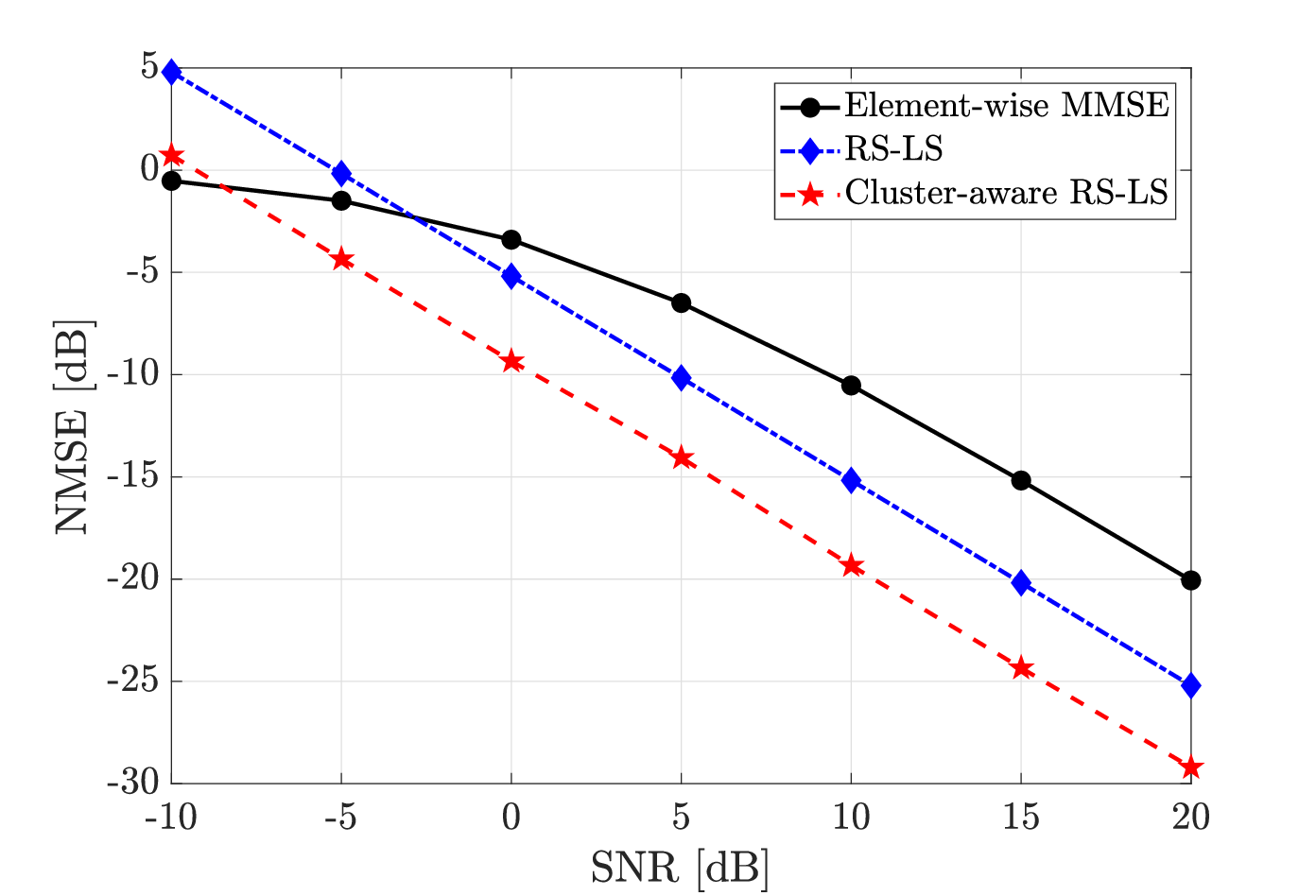}
			\vspace{-0.2cm}
			\caption{NMSE versus SNR for different estimators  with $\tau_p=K=64$.} \label{fig:RSLS-taup-K} \vspace{-4mm}
\end{figure}

\begin{figure}[t!]
\hspace{0mm} 
\includegraphics[trim={0.7cm 0.1cm 1.5cm 0.2cm},clip,width=3.2in]{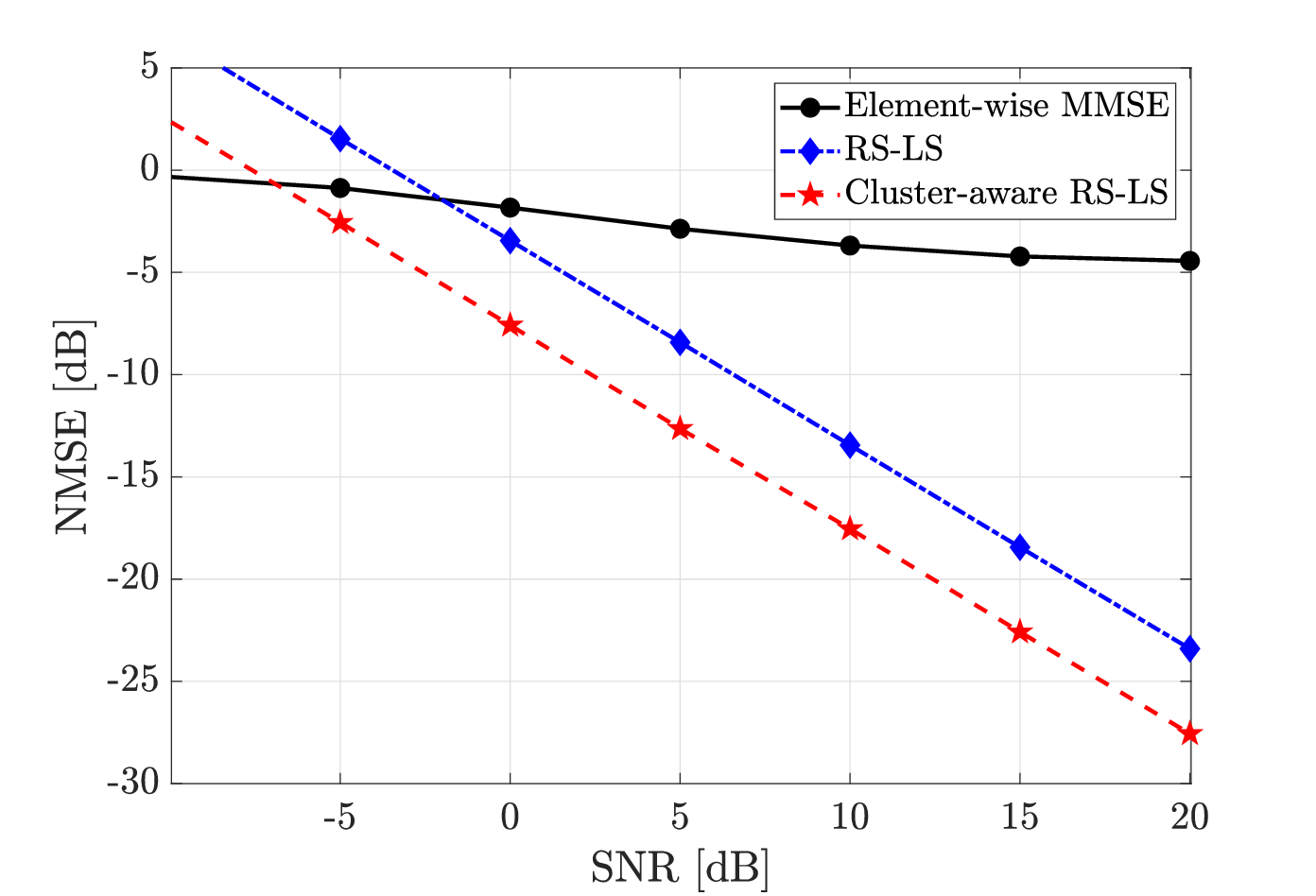}
			\vspace{-0.2cm}
			\caption{NMSE versus SNR for different estimators with $\tau_p=43$.} \label{fig:RSLS-taup-short} \vspace{-4mm}
\end{figure}

In Figs. \ref{fig:RSLS-taup-K} and \ref{fig:RSLS-taup-short}, each point corresponds to the average outcome derived from 25 random channel realizations. The vectorized channel $\mathbf{x}$ is generated in accordance with \eqref{eq:clustered-channel}, where each $\mathbf{w}_{\ell}$ contains independent and identically distributed complex Gaussian entries. The channel is normalized such that $\tr(\vect{R}) = MK$. On the horizontal axis of the figures, the signal-to-noise ratio (SNR) represents the ratio of the product of channel gain and transmit power to the receiver noise variance.

In Fig.~\ref{fig:RSLS-taup-K}, we fix the pilot length at $\tau_p=K=64$ and vary the SNR to evaluate the NMSE performance of the three channel estimators. As illustrated, except at very low SNR levels, the RS-LS estimator significantly outperforms the element-wise MMSE estimator, achieving approximately a 5\,dB improvement in NMSE. This enhancement stems from the noise rejection in the unused dimensions, which constitute the null-space of the subspace where any channel may reside. When we employ the newly proposed cluster-aware RS-LS scheme, an additional improvement of 4\,dB is noted. Importantly, the cluster-aware RS-LS estimator leverages the structured subspace introduced by clustered multipath propagation without requiring knowledge of the spatial correlation matrix. Neither of the RS-LS methods utilizes specific entries from the spatial correlation matrix.

In Fig.~\ref{fig:RSLS-taup-short}, we decrease the pilot length to $\tau_p=43$, where $43$ represents the effective rank of $\vect{R}_{\rm t,iso}$, capturing a fraction $1-10^{-5}$ of the total eigenvalue sum. The proposed RS-LS estimator maintains satisfactory channel estimation accuracy even when $\tau_p$ is less than $K$. Moreover, as SNR increases, the performance disparity between the element-wise MMSE estimator and the RS-LS estimators becomes more pronounced. As illustrated, the reduced pilot length inhibits NMSE reduction in the element-wise MMSE estimator.

\section{Conclusions}
Channel estimation in a extremely large-scale MIMO is a complex problem since acquiring all the entries of the spatial correlation matrix is practically challenging.
In this paper, we extend the previously developed RS-LS estimator from single-input multiple-output (SIMO) channels to the MIMO context and introduce two innovative channel estimators that exploit the reduced-dimension subspace induced by array geometry. This reduced dimensionality is an inherent characteristic of UPAs, arising from non-existent channel dimensions that primarily carry noise. By eliminating noise from these dimensions, it is possible not only to enhance the performance of the classical LS and element-wise MMSE estimators  but also to reduce the required pilot length owing to the reduced dimensions. Beyond the conventional RS-LS estimator developed using our established frameworks, we have also crafted a novel channel estimator that utilizes orthogonal transmit subspaces and their corresponding receive subspaces for clustered multipath propagation. The RS-LS approach enables a decrease in NMSE by 5\,dB compared to the element-wise MMSE with the conventional RS-LS, and an additional 4\,dB reduction with the new cluster-aware RS-LS estimator.

\bibliographystyle{IEEEtran}
\bibliography{IEEEabrv,refs}

\end{document}